\documentclass[12pt, a4paper]{article}
\usepackage[margin={2cm,2cm}]{geometry}
\pdfoutput=1
\usepackage{cite}
\usepackage{siunitx}
\usepackage{amsmath}
\usepackage[breaklinks=true,colorlinks=true,linkcolor=blue,urlcolor=blue,citecolor=blue]{hyperref}
\usepackage{graphicx}
\usepackage[font=small,labelfont=bf]{caption}
\usepackage{subcaption}
\usepackage{gensymb}
\usepackage{authblk}
\usepackage{soul}
\usepackage{xfrac}
\usepackage[switch]{lineno}
\usepackage{xfrac}
\usepackage{float}

\title{Structure-stiffness relation of live mouse brain tissue determined by depth-controlled indentation mapping}
\author[1,*]{Nelda Antonovaite}
\author[1]{Steven V. Beekmans}
\author[2,3,4]{Elly M. Hol}
\author[4]{Wytse J. Wadman}
\author[1]{Davide Iannuzzi} 

\affil[1]{\small{Department of Physics and Astronomy and LaserLab Amsterdam, Vrije Universiteit Amsterdam, De Boelelaan 1085, 1081 HV, Amsterdam, Netherlands, nelda.antonovaite@student.vu.nl}}
\affil[2]{\small{Department of Translational Neuroscience, Brain Center Rudolf Magnus, University Medical Center Utrecht, 3584 CG Utrecht, The Netherlands}}
\affil[3]{\small{Netherlands Institute for Neuroscience, An Institute of the Royal Netherlands Academy of Arts and Sciences, 1105 BA Amsterdam, The Netherlands}}
\affil[4]{\small{Center for Neuroscience, Swammerdam Institute for Life Sciences, University of Amsterdam, 1098 XH Amsterdam, The Netherlands}}

\begin{document}
		\maketitle
		
		\begin{abstract} 
			\noindent The mechanical properties of brain tissue play a pivotal role in neurodevelopment and neurological disorders. Yet, at present, there is no consensus on how the different structural parts of the tissue contribute to its stiffness variations. Here, we have gathered depth-controlled indentation viscoelasticity maps of the hippocampus of isolated horizontal live mouse brain sections. Our results confirm the highly viscoelestic nature of the material and clearly show that the mechanical properties correlate with the different morphological layers of the samples investigated. Interestingly, the relative cell nuclei area seems to negatively correlate with the stiffness observed.\\
		\end{abstract}
	
	
Brain tissue consists of neuronal cell bodies, their processes (dendrites and axons, myelinated or not, which form either sparse branches and arborizations or dense fiber bundles), the
interconnecting extracellular brain matrix (ECM), glial cells, blood vessels, and extracellular fluid. Each of these components may have a different influence on the local
mechanical properties of the tissue, which, in turn, regulate a wide variety of very relevant mechanotransduction phenomena. For instance, mechanical signals are known to play a role
in multiple vital processes of neural cells~\cite{moshayedi_relationship_2014, jagielska_mechanical_2012}, whereas neuronal growth, neurite extension, arborization patterns, neuronal
traction forces, and the stiffness of individual neurons and glial cells were all found susceptible to the stiffness of the substrate~\cite{koser_mechanosensing_2016,
moshayedi_mechanosensitivity_2010, cullen_collagen-dependent_2007, georges_matrices_2006, teixeira_promotion_2009, sur_tuning_2013, flanagan_neurite_2002, bollmann_microglia_2015,
chen_statistical_2016}. Furthermore, an abnormal mechanical environment (emerging as a result of internal or external forces, changes in the
composition of the ECM, or changes in osmotic conditions) can disrupt normal brain function and neurodevelopment, and can alter progression of neurological
disorders~\cite{kumar_mechanics_2009, barros_extracellular_2011, vegh_reducing_2014}. It is therefore commonly agreed that a deep knowledge of the correlation between brain composition
and mechanical properties of the tissue would enable neuroscientists to shed light on how mechanotransduction phenomena contribute to the functioning of the brain. Furthermore, a
quantitative assessment of the viscoelasticity characteristics of the different regions of the brain could pave the way for the improvement of computational brain injury
models~\cite{mao_why_2013}, the engineering of brain phantoms for surgical practise~\cite{forte_composite_2016,Chen_anthropomorphic_2012}, the design of mechanically matched brain
implants~\cite{spencer_characterization_2017}, and the production of soft substrates that could mimic different mechanical environments for investigations of stiffness-dependent neural stem cell
differentiation~\cite{saha_substrate_2008, keung_pan-neuronal_2013,luque_microelastic_2016} and neuronal and glial cell morphology~\cite{Puschmann_bioactive_2013}. Yet, at present, the
relation between mechanical properties and cytoarchitecture is still largely unknown.

Previous studies on brain mechanics have been mainly limited to the comparison between white and gray matter, with results that are either inconsistent or lack quantitative structure
analysis~\cite{budday_mechanical_2015, weickenmeier_brain_2016, van_dommelen_mechanical_2010, kaster_measurement_2011, feng_measurements_2013, forte_characterization_2017,
christ_mechanical_2010, koser_cns_2015, samadi-dooki_indirect_2017}. Only recently, one study has shown that the mechanical properties correlate with myelin content in bovine
brain~\cite{weickenmeier_brain_2016}, while stiffness was found to scale with the cell nucleus area in spinal cord of mouse and retinal ruminant tissue~\cite{koser_cns_2015,
weber_role_2017}. Even though differences in mechanical properties of large hippocampal regions such as cornu ammonis (CA) fields and the dentate gyrus (DG) have been reported
previously~\cite{finan_viscoelastic_2012,elkin_dynamic_2011,elkin_viscoelastic_2013,elkin_age-dependent_2010,finan_viscoelastic_2012_1}, this set of data is not sufficient to
completely determine the correlation between the morphological structure of the brain and its mechanical properties.

From a purely technical perspective, this literature gap is not surprising. Brain tissue is a highly viscoelastic, non-linear, anisotropic material~\cite{franceschini_brain_2006,
hrapko_influence_2008,feng_measurements_2013,budday_mechanical_2017,labus_anisotropic_2016,chatelin_fifty_2010}, and, because of its cellular heterogeneity, low stiffness, and rapid degradation, it is one of the most difficult (bio)materials to mechanically test. Macroscopic tests (such as shear rheology, compression testing, and tension testing) can only measure the mechanical properties of large samples, and, for this reason, cannot provide information on the local features of the tissue. Atomic force microscope (AFM) indentation, on the contrary, makes use of a small radius tip to locally probe the mechanical response of a material to a compressive stress, and, therefore, seems to be more suitable to assess how the mechanical properties of the different regions of the brain may be influenced by the underlying morphological composition~\cite{Elkin_mechanical_2007}. Unfortunately, the results reported in the literature do not always agree with each other, as witnessed by the wide range of stiffness values reported~\cite{chatelin_fifty_2010,hrapko_influence_2008,cheng_rheological_2008}.

Tantalized by this challenge, we have explored whether a recently introduced mechanical testing technique, known as ferrule-top dynamic indentation~\cite{chavan_ferrule-top_2012,hoorn_local_2016}, could provide a better insight on the correlation between the composition of brain tissue and its viscoelastic properties. The technique used in our experiment is quite similar to AFM indentation. However, with respect to the latter, it offers one key advantages, in that the position of the cantilever is monitored by means of optical fiber interferometry rather than the optical triangulation technique used in AFM. As already showed in several papers~\cite{chavan_ferrule-top_2012,hoorn_local_2016,beekmans_minimally_2017,beekmans_characterizing_2016}, this method is suited for the implementation of highly stable electromechanical feedback loops, which, in turn, guarantee better measurement protocols, including dynamic mechanical analysis at controlled indentation depth. 

In this paper, we demonstrate that, using our deep, depth-controlled indentation method, one can obtain high spatial resolution viscoelasticity map of a mouse hippocampus slice. The map clearly emphasizes how the different structural layers give rise to different mechanical properties. Our results further show that brain tissue appears stiffer as the indentation depth or frequency increase -- a result that confirms the non-linear viscoelastic nature of the brain tissue. Finally, calculating the mean measured stiffness of eleven anatomical subregions, and comparing it with the estimated nuclear densities, we can infer that densely packed cell layers may actually have lower stiffness than more disperse ones -- a result that seems to contradict the commonly accepted assumption that brain tissue stiffness is dominated by cell bodies~\cite{koser_cns_2015,koser_mechanosensing_2016}.	
	
\section*{Results}
\subsection*{Mechanical heterogeneity of hippocampus agrees with anatomical region boundaries}
\begin{figure}[H]
	\centering
	\includegraphics[scale=0.7]{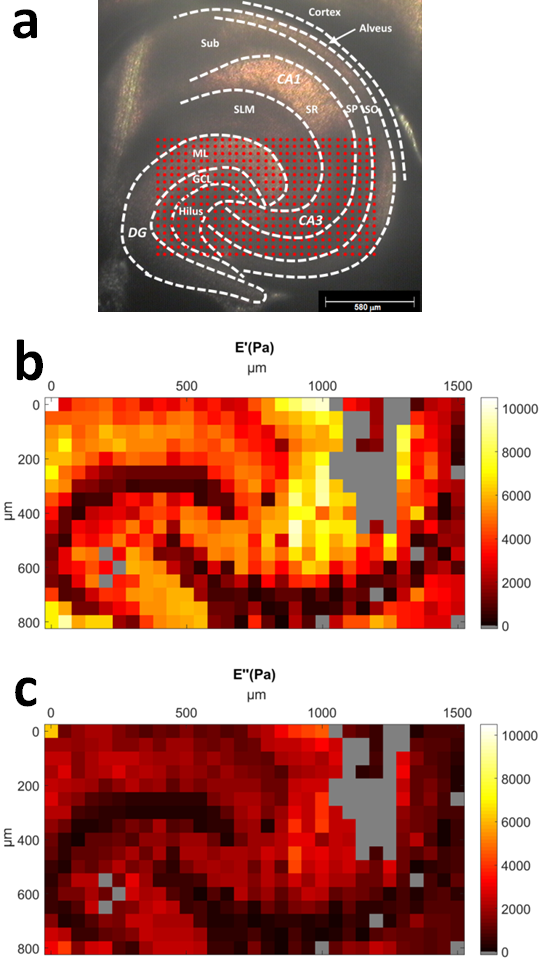}
	\caption{\label{modulus} (a) Microscope image of one of the slices used in this experiment, along with a typical 50~\SI{}{\micro\meter} $\times$ 50~\SI{}{\micro\meter} indentation grid (red dots). Dashed lines indicate boundaries of morphologically distinct anatomical regions. Abbreviations: Sub - subiculum, SLM - stratum lacunosum moleculare (SLM), SR - stratum radiatum, SP - stratum pyramidale, SO - stratum oriens, ML - molecular layer, GCL - granule cell layer; large regions: DG - dentate gyrus, CA - cornus ammonis. (b, c) Color-coded map of storage \textit{E}\textasciiacute (b) and loss modulus \textit{E}\textacutedbl (c) in Pa over the DG and CA3 field of
		the hippocampal formation obtained with oscillatory ramp depth-controlled indentation strokes (see Online Methods) with 0.2~\SI{}{\micro\meter} oscillation amplitude, 5.62~Hz oscillation frequency, and 9$\%$ strain. Gray color indicates failed measurements.}
\end{figure}

Viscoelasticity maps of live brain sections were obtained by means of ferrule-top depth-controlled indentation~\cite{chavan_ferrule-top_2012,hoorn_local_2016}. The image of one of the samples used in the experiment, along with a typical 50~\SI{}{\micro\meter} $\times$ 50~\SI{}{\micro\meter} grid of indentation locations, can be seen in Fig.~\ref{modulus}a. We refer the reader to the method section for further details on instrumentation and protocol.
Fig.~\ref{modulus}b and c show the viscoelasticity maps (\textit{E}\textasciiacute = storage modulus; \textit{E}\textacutedbl = loss modulus) over the DG and the proximal portion of CA3 field of the hippocampus obtained from a horizontal mouse (9 months old) brain slice around 3 to 4~mm in the dorsal-ventral position. The representative maps were obtained with depth-controlled oscillatory ramp indentation strokes (see Online Methods) at an estimated strain of 9$\%$. Similar
maps, albeit focused on smaller areas, were obtained in 8 other slices (7 slices from 6 months old mice and 1 slice from 9 months old mice) out of 11 tested. As for the remaining 2 (both obtained from 9 months old mice), the data looked rather scrambled and not reproducible, probably because the sample was not perfectly adhering on the substrate.

In Fig.~\ref{modulus}, one can clearly identify multiple areas with distinctive mechanical features. The shape of these regions agrees well with the morphological heterogeneity
of anatomical subregions (see Fig.~\ref{modulus}a), including the U-shaped structure of the molecular layer (ML) and of the granule cell layer (GCL) enclosing the hilus, and the laminar organization of strata (layers): oriens (SO), pyramidale
(SP), radiatum (SR) and lacunosum-moleculare (SLM).
\subsection*{Brain tissue is non-linear viscoelastic}
\begin{figure}[H]
	\centering
	\includegraphics[scale=0.55]{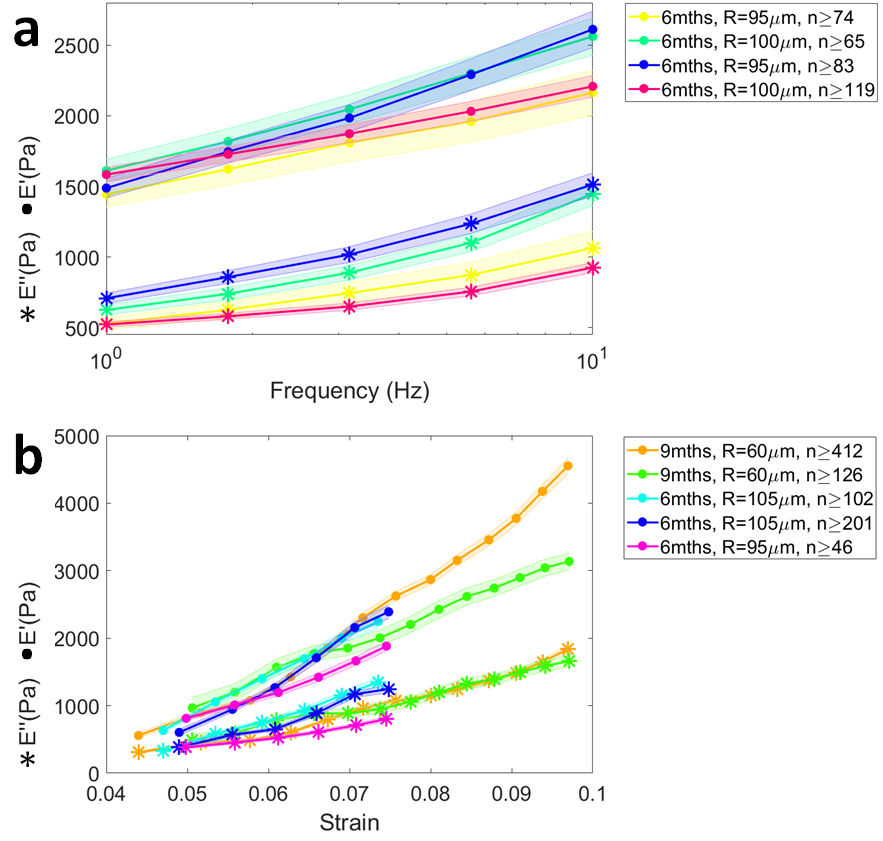}
	\caption{\label{frequency} Non-linear viscoelastic properties of hippocampus obtained with dynamic indentation-controlled testing: frequency sweep (a) and oscillatory ramp (b). (a) Storage and loss moduli \textit{E}\textasciiacute~and \textit{E}\textacutedbl~ increase over a frequency range of 1-10~Hz (data averaged over hippocampus;
		measured at $\sim$7$\%$ strain, 0.2~\SI{}{\micro\meter} oscillation amplitude; note the logarithmic scale on the x-axis). (b) \textit{E}\textasciiacute~and \textit{E}\textacutedbl~increase over the strain range of 4-10$\%$ (measured at 5.62~Hz oscillation frequency, 0.2~\SI{}{\micro\meter} oscillation amplitude). The age of the animals is specified in the legend. Shadowed zones indicate the standard error of the mean, whereas $R$ indicates the radius of the indenting tip and $n$ the number of measurement points.}
\end{figure}

To confirm that ferrule-top depth-controlled dynamic indentation is capable of capturing the non-linear viscoelastic nature of the brain tissue, we analyzed all the data obtained from the hippocampus together
(i.e., without subdivision into layers). The averaged storage and loss moduli over frequency (obtained with the frequency sweep method) and strain (obtained with the oscillatory ramp method) are shown in Fig.~\ref{frequency}. The frequency sweep data reveal a stiffening of the tissue with increasing indentation frequency, whereas the depth profiles from the oscillatory ramp testing show a stiffening with strain.

\subsection*{Regional viscoelastic properties are reproducible}

\begin{figure}[H]
	\centering
	\includegraphics[scale=0.60]{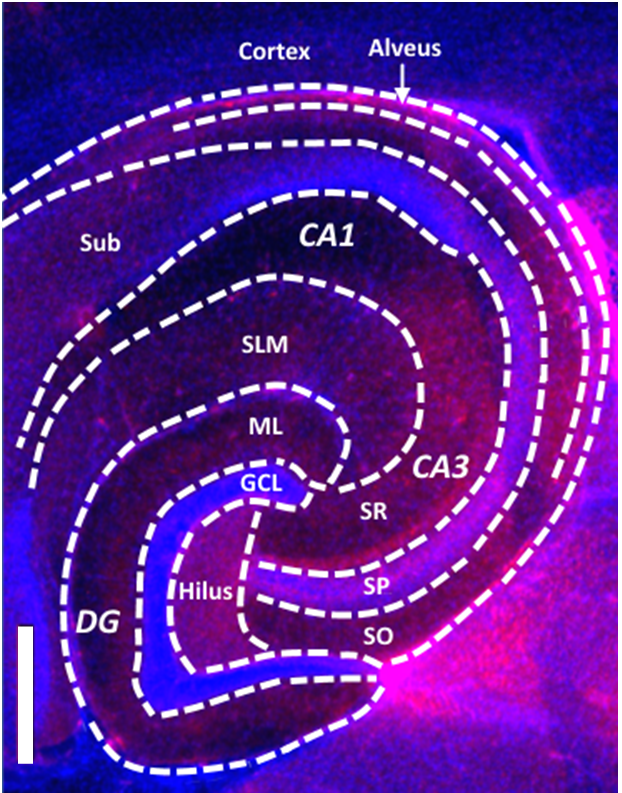}
	\caption{\label{anatomy} Fluorescent microscopy image of the hippocampus, visualizing cell nuclei (blue) and neurofilaments (red). Dashed lines indicate boundaries of morphologically distinct anatomical regions. Scale bar = \SI{200}{\micro\meter}.}
\end{figure}

To test inter-animal variability of the mechanical properties measured in different brain areas, we developed a protocol to identify anatomical regions and indentation locations. The
coordinates of the XYZ micromanipulator were calibrated, prior to the measurements, using the image of the tip of the probe in the camera of the inverted microscope. At the end of the indentation measurement, each measured brain section was formalin-fixed and stained for nuclei and neurofilaments, which indicates the neuronal axons (see Fig.~\ref{anatomy}, Online Methods). Differences in cell densities and organization of axons, visualized in fluorescent images, allowed us to draw boundaries between anatomical regions and overlay them with the image of the slice from the inverted microscope. Next, each indentation location was assigned to the corresponding anatomical region and the viscoelastic properties were averaged over these regions. The layered composition of the cortex areas varied between slices and were treated as a single region.

\begin{figure}[H]
	\centering	
	\includegraphics[width=.8\linewidth]{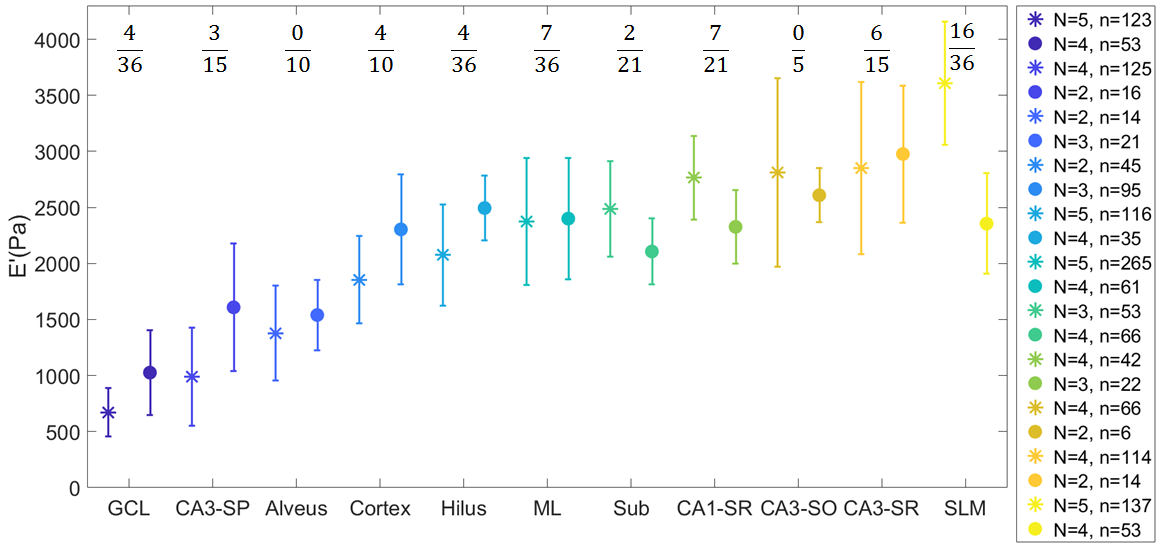}
	\caption{\label{order} Comparison of the weighted means of the storage modulus of different brain regions
		obtained with an oscillatory ramp (stars) or a frequency sweep (dots) approach at 7.3$\%$ strain and 5.62~Hz frequency. The legend indicates the number of slices $N$ used to calculate the weighted mean and the total number of
		measurement points $n$. The fraction on top indicates the number of significantly different pairs over the total number of pairs analyzed with \textit{post hoc} test. Error bars are SE of the weighted mean.}
	\label{bp}
\end{figure}

Fig.~\ref{order} shows the value of \textit{E}\textasciiacute at 7.3$\%$ strain and 5.62~Hz frequency, averaged over all the slices, for each of the different regions identified with the staining procedure, plotted from the softest to the stiffest in increasing stiffness order (as determined by the results obtained with the oscillatory ramp method). As expected from the viscoelasticity maps, the mechanical properties of the brain tissue appear highly heterogeneous. Both measuring methods (oscillatory ramp and frequency sweep) highlight the same trend in the mechanical properties of the different regions investigated, with the exception of the SLM region, where the frequency sweep method seems to indicate a decrease in stiffness that the oscillatory ramp does not detect.

To compare the local storage modulus measured in different
slices, we performed one-way ANOVA followed by \textit{post hoc} test for each region (Online Methods). The results are indicated on top of Fig.~\ref{order} as a ratio between the number of significantly different pairs over the total number of pairs used for the comparison. Remarkably, only 17$\%$ of the tested pairs were significantly different, especially if one considers that 57$\%$ of these variations stems from the comparison of data obtained with different experimental methods (frequency sweep versus oscillatory ramp). This result confirms that, in our experiments, there is a good inter-animal reproducibility of the results.

\subsection*{Relative area of nuclei negatively correlates with mechanical properties}

\begin{figure*}
	\centering	
	{\includegraphics[width=.8\linewidth]{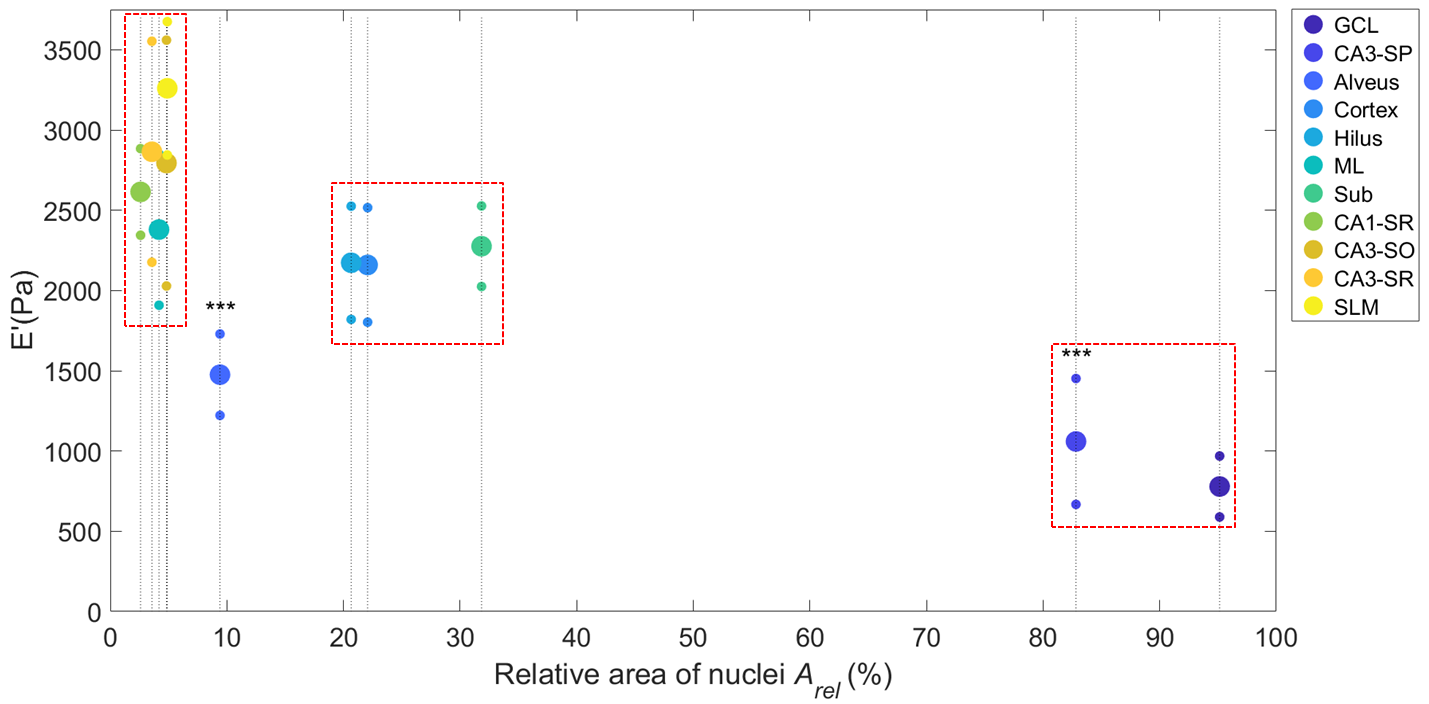} }
	\caption{\label{densities}
		Weighted means of the storage modulus (combined data from all slices) plotted as a function of relative area of nuclei. Stars indicate regions with high axon density. Red dashed squares mark three groups based on stiffness and $A$$_{rel}$ (from left to right): stiff and low-$A$$_{rel}$,
		intermediate stiffness-$A$$_{rel}$, soft and high-$A$$_{rel}$. Error bars are SE of the weighted mean.}	
	\label{bp}
\end{figure*}

The relation between stiffness and underlying morphological structure was assessed by evaluating the percentage of area covered by cell nuclei of each of the regions investigated. A horizontal Nissl-stained slice from a similar location within the brain was used to obtain relative nuclei area $A$$_{rel}$ in each measured region (see Online Methods). Afterwards, the weighted mean of the storage modulus of all slices was plotted against $A$$_{rel}$ (Fig.~\ref{densities}). In this graph, one can identify three groups. The first group from the left, which includes the SR of the CA1 and the CA3 fields, the SO-CA3 and the SLM, appears to be the stiffest ($\sim$2614-3260~Pa) and relatively cell-free ($A$$_{rel}$ = 2.6-4.9$\%$). The second group, which includes the cortex, the hilus and the subiculum
has intermediate stiffness ($\sim$2159-2276~Pa) and intermediate $A$$_{rel}$ (20.7-31.8$\%$). Finally, the GCL, which has the highest $A$$_{rel}$ (95.1$\%$), also appears to have the softest mechanical properties ($\sim$779~Pa), followed by the SP-CA3 with an $A$$_{rel}$ of 82.8$\%$
($\sim$1059~Pa). While the SP-CA3 layer has a high density of both axons and cells, we could not evaluate the contribution of axons to the stiffness. However, alveus, containing mostly
fibers and low $A$$_{rel}$ (9.4$\%$) appears to be softer ($\sim$1475~Pa) than all other high-intermediate stiffness and low-intermediate $A$$_{rel}$, suggesting that high density of fibers decreases the stiffness of the tissue in low-intermediate $A$$_{rel}$ regions.

Remarkably, by carrying out a linear regression on all data of Fig.~\ref{densities}, one can show that there is a strong negative correlation between the stiffness of a brain region and its relative area of nuclei (Pearson's correlation coefficient $r$ = -0.85), which is even stronger if the Alveus is excluded from the analysis ($r$ = -0.97). 

\begin{figure*}
	\centering	
	{\includegraphics[scale=0.40]{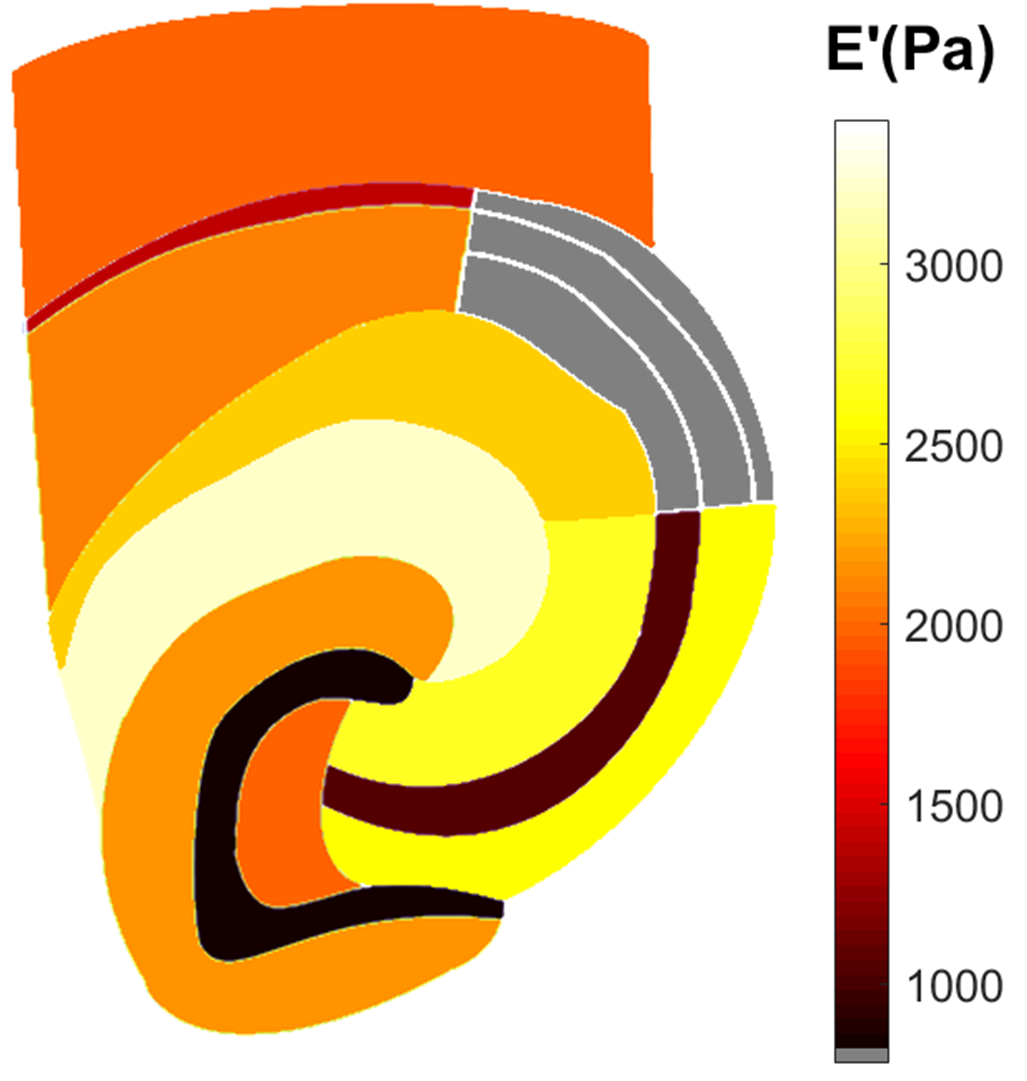} }
	\caption{\label{reconstruction}
Reconstructed storage modulus map of brain tissue regions based on a weighted mean of 9 slices. Gray color indicates regions that were not measured.}
\label{bp}
\end{figure*}

For future reference, in Fig.~\ref{reconstruction}, we provide a map of the storage modulus of the different brain tissue regions as reconstructed with the values reported in Fig.~\ref{densities}.

\section*{Discussion}

 In this study, we have used a ferrule-top indentation approach to gather viscoelastic maps of mouse brain tissue in their physiological state. The size of
the indentation sphere and the depth of indentation were selected to ensure that the measurements could provide the tissue mechanical features of the subregional area of the tested sample. High spatial resolution (\SI{50}{\micro\meter}) of indentation mapping allowed us to find a clear correlation between indented regions and viscoelastic properties. The same relationship was reproduced on multiple slices by means of different testing method (frequency sweep and oscillatory ramp). 

 Our measurements show that both storage and loss moduli increase with strain. We can thus confirm that brain tissue is a non-linear material, as already reported in other studies~\cite{elkin_age-dependent_2010, elkin_dynamic_2011, koser_cns_2015}. Furthermore, performing the first localized frequency-domain indentation measurements on brain slices ever reported in the literature, we could observe that, in the 1-10~Hz range, both storage and loss moduli increase with indentation frequency -- a viscoelastic behavior that was previously observed as a stress relaxation, creep response and change in strain-rate sensitivity in other indentation experiments~\cite{budday_mechanical_2015,van_dommelen_mechanical_2010,elkin_dynamic_2011,elkin_viscoelastic_2013,finan_regional_2017}. In quantitative terms, it is interesting to note that, converting the averaged values of $E$\textasciiacute~obtained in our measurements~(Fig.~\ref{frequency}a) to shear moduli $G$\textasciiacute~by dividing $E$\textasciiacute by 2(1+$\nu$) (where $\nu=0.5$ is the Poisson's ratio of compressibility), one obtains values for $G$\textasciiacute~of 0.5-0.8$\pm$0.1~kPa, which is in good agreement with macroscopic (i.e., not localized) frequency sweep measurements reported in the literature~\cite{brands_large_2000, hrapko_mechanical_2006, peters_applicability_1997, shen_modified_2006,samadi-dooki_indirect_2017,chatelin_towards_2012, forte_characterization_2017, garo_towards_2007, vappou_magnetic_2007}. A direct comparison with other local measurements is unfortunately not possible, because the latter have been either performed on different kind of samples (different age, species, slicing direction) or under very diverse indentation stroke protocols, which, because of the highly viscoelastic behavior of the material, give rise to very different results~\cite{finan_viscoelastic_2012,elkin_viscoelastic_2013,elkin_age-dependent_2010,elkin_dynamic_2011,christ_mechanical_2010,prange_regional_2002,weickenmeier_brain_2016,budday_mechanical_2015,van_dommelen_mechanical_2010,kaster_measurement_2011}.

Our viscoelasticity maps further reveal that the stiffness of the hippocampal subregions and the cortex negatively correlates with the relative area of nuclei; in other words, higher cell body density corresponds to a softer tissue. This is consistent with the study on single cortical neurons were soma has been found to be significantly softer than 
neurites~\cite{grevesse_opposite_2015}. However, our findings are in contrast with previous studies, which found a positive correlation between stiffness and cell nucleus area on the spinal cord of the adult mouse, retina of the ruminant and embryonic brain of \textit{Xenopus}~\cite{koser_cns_2015,koser_mechanosensing_2016,weber_role_2017}. Furthermore, granular cell layer in coronal hippocampal slices of the juvenile rats was also shown to be stiffer than hilus~\cite{luque_microelastic_2016}. While different CNS tissue composition might be the reason for the discrepancy between our results and those reported in the literature, it is worth stressing that our indentation protocol significantly differs with respect to that used in previous studies. The AFM system used in previous studies relies on small beads (radius $<$ \SI{20}{\micro\meter}) and on a stroke that moves the probe at a constant speed of 10~\SI{}{\micro\meter}/s until a predifined maximum force is reached (with the maximum force being on the order of several nN). We estimate that the contact radius must then be smaller than \SI{10}{\micro\meter}, with an indentation depth of less than \SI{4}{\micro\meter}. It is thus legitimate to ask whether this kind of AFM measurements probe the tissue properties or, actually, only indent the first layer of cells that lie on the surface. Furthermore, the piezo-control testing used in AFM measurements results in different strain rates and indentation depth for different values of stiffness of the tissue indented; under the same stroke protocol, a softer tissue will experience a higher strain rate and a larger indentation depth. In contrast, our deep, indentation-control testing protocol assures constant indentation depth and constant indentation speed. Furthermore, with indentation beads in the range of 60-105~\SI{}{\micro\meter}, strain of $\sim$7$\%$, and indentation depth between 8-11~\SI{}{\micro\meter} (which results in a contact radius between 20 and 40~\SI{}{\micro\meter}), we are sure to measures at the scale of the tissue. Therefore, we suggest that the opposite relationship between stiffness and areas of cell nuclei observed in our experiment might be at least partially due to the difference in the scale probed and/or in the very same testing method.

 Our data is not sufficient to fully explain why low nuclear density regions translate into stiffer tissues. At first, one may think that the increase of stiffness with the decrease of nuclear density be due to the role of the perineural nets (PNNs). However, it has been showed that this component lacks fibrous proteins~\cite{bonneh-barkay_brain_2009}, and, therefore, should not support mechanical loading. Another component that may play a role in the mechanical properties of the brain tissue is myelin, as it has been shown that brain stiffness increases with myelin content~\cite{weickenmeier_brain_2016, elkin_age-dependent_2010}. Yet, our results indicate that the bundle of myelinated fibers in the alveus and the tract of mossy fibers along the SP in the CA3 field are actually softer. One may thus speculate whether the mechanical properties of the brain tissue as observed in our experiment are rather due to the fact that regions with low nuclear density host a larger amount of sparsely distributed axons and dendrites under tension~\cite{essen_tension-based_1997,
	haslach_solidextracellular_2014,franze_mechanical_2013,franze_biophysics_2010,xu_residual_2009,xu_axons_2010,hilgetag_role_2006}, which may in turn give rise to a stiffer material. To confirm or reject this hypothesis, it is imperative to perform new measurements that could directly correlate stiffness with axon density and orientation.

{\section*{Methods}
{\subsection*{Sample preparation}
	\label{sample}

	C57Bl/6 mice were sacrificed at an age of 6 or 9 months. All experiments were performed in accordance with protocols and guidelines approved by the Institutional Animal Care and Use
	Committee (UvA-DEC) operating under standards set by EU Directive 2010/63/EU. All efforts were made to minimize the suffering and number of animals. The mice were decapitated, the
	brain was removed from the skull and stored in ice-cold ACSF containing (in mM): 120 NaCl, 3.5 KCl, 5 MgSO$_{4}$, 1.25 NaH$_{2}$PO$_{4}$, 2.5 CaCl$_{2}$, 25 NaHCO$_{3}$, and 10 glucose
	oxygenated with 95$\%$ O$_{2}$/5$\%$ CO$_{2}$ ($\sim$310~mOsmol/kg and $\sim$pH 7.4). Slices were cut in a horizontal plane with a thickness of approximately 300~\SI{}{\micro\meter} using a VT1200S
	vibratome (Leica Biosystems, Nussloch, Germany). Afterwards, a single brain tissue slice was placed in a perfusion chamber maintained at $\sim$20~$\degree$C and supplied with carbogen
	saturated ACSF solution at 1~ml/min flow rate (gear pump MCP-Z standard, Ismatec). The glass bottom of the chamber was coated with 0.05$\%$ polyethylenimine for the adhesion with the
	sample, which was also pressed down from the top with a 2~mm spaced harp to ensure the stability. After acclimatization for 1~h, indentation measurements were performed within 8~h.
	
	{\subsection*{Imaging and immunfluorescence}
	\label{immuno}
	An inverted microscope (Nikon TMD-Diaphot, Nikon Corporation, Japan) was used to image the slice during the measurements with a 2 $\times$ magnification objective (Nikon Plan 2X, Nikon
	Corporation, Japan). Images were recorded with a CCD camera (WAT-202B, Watec). After the measurements, slices were fixed in 4$\%$ PFA overnight at 4$~\degree$C. The sections were
	washed 3$\times$10~min in PBS solution (0.01~M, pH = 7.4) and subsequently blocked for 2~h with 10$\%$ normal donkey serum (NDS), 0.25$\%$ Triton in PBS solution. After, slices were
	incubated overnight at 4$~\degree$C with primary antibody against neurofilament 160~kDa (NN18 N5264, Sigma Aldrich, 1:1000 dilution), 3$\%$ NDS and 0.25$\%$ Triton in PBS solution.
	Consequently, slices were washed 3$\times$10~min in PBS solution and incubated for 2~h in PBS solution with DNA stain (Hoechst 33342, Invitrogen, dilution 1:2000), the secondary
	antibody Cy3-conjugated donkey anti-mouse IgG (Jackson Immuno Research, 1:1400 dilution), 1$\%$ NDS and 0.025$\%$ Triton. Finally, sections were washed 3$\times$10~min in PBS and
	mounted with a glass coverslip in Vectashield (Vector Laboratories). Fluorescent images were obtained with Leica DMRE fluorescence microscope (Leica Microsystems, Wetzlar, Germany) and overlaid with the previously obtained
	bright-field images to identify anatomical regions of measured locations. Immunofluorescently-labeled slices were too thick for objective calculation of cell density and, thus, percentage of area covered by nuclei
	was estimated for each region from a Nissl stained image of horizontal 25~\SI{}{\micro\meter} thickness adult-mouse brain section (http://brainmaps.org,~\cite{Mikula_internet-enabled_2007}):
	\begin{align}
		A_{rel}	=\dfrac{A_{nuclei}}{A_{total}},
	\end{align}
	
	\noindent where $A$$_{nuclei}$ is the area covered by nuclei in the region and $A$$_{total}$ is the total area of that region (processed with image J).
	
	\subsection*{Dynamic indentation setup and measurement protocol}
	\label{protocol}
		\begin{figure}[H]
		\centering
		\includegraphics[scale=0.40]{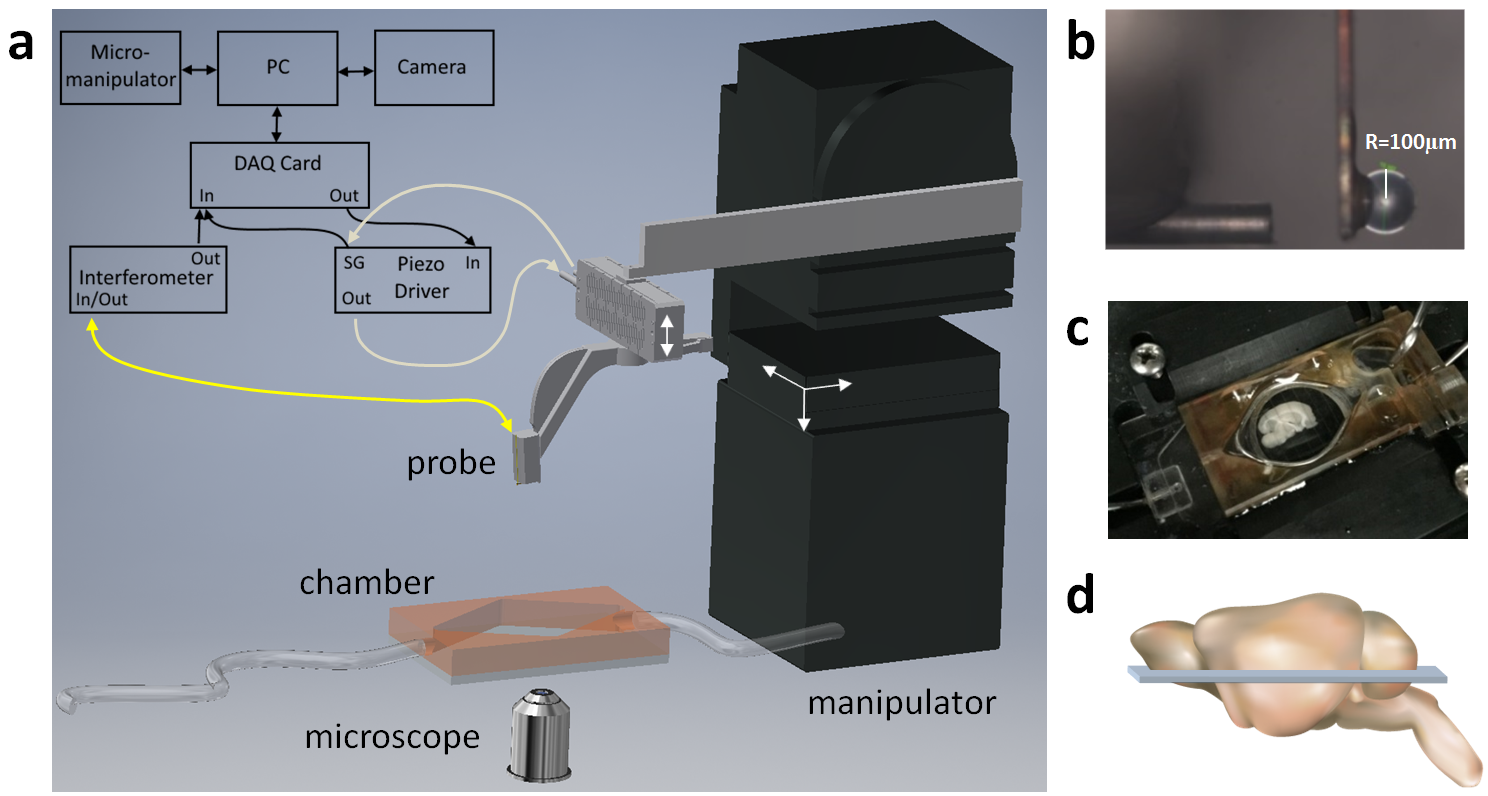}
		\caption{\label{setup}Schematic view of the dynamic indentation mapping setup. A ferrule-top probe (a) is equipped with an optical fiber for interferometric readout of the
			cantilever position and with a spherical tip to indent the sample (b). The probe is mounted on a piezoelectric transducer (a) for controlled movement during an indentation
			measurement. A brain slice is submerged in the perfusion chamber and fixed with the harp (c). The slice is imaged with an inverted microscope (a) for the determination of indentation locations. (d) The approximate position of the slice within the brain.}
	\end{figure}
	Horizontal mouse brain slices from 3 to 4~mm of dorsal-ventral positions of hippocampus (Fig.~\ref{setup}d) were submerged in a perfusion chamber assembled on the microscope, pressed
	down with a 2~mm spaced harp to assure stability, and supplied with a constant flow of carbogenated artificial cerebrospinal fluid (ACSF). Measurements were carried out on 9 slices
	from eight mice of which 6 were 6 months old and 2 -- 9 months old. The indentation lines were selected to cross the DG and the subiculum or the CA3 field of hippocampus ($n$ $\geq$ 66 measurement points per slice). In
	addition, indentations on cortex were performed on 5 of the same slices adjacent to subiculum, in parallel lines between outer and inner layers ($n$ $\geq$ 21 measurement points per
	slice).
	
	A ferrule-top force transducer~\cite{chavan_ferrule-top_2012}, consisting of a micromachined cantilever spring with optical fiber readout, was mounted on a 3D printed holder screwed to a Z-piezoelectric actuator (PI p-603.5S2, Physik Instrumente). The single-mode fiber of the readout was coupled to an interferometer (OP1550, Optics11), where the interference signal was directly translated into cantilever deflection. Indentation depth control was implemented through a feedback loop, based on the error signal of cantilever deflection (Fig.~\ref{setup}, for more details see~\cite{hoorn_local_2016}). The piezoelectric actuator with the probe was mounted on a XYZ micromanipulator (PatchStar, Scientifica) for automatic mapping of mechanical properties. Indentation mapping was performed in parallel lines, with 59-476 points per slice. Distance between two adjacent locations were in the range of 50-160~\SI{}{\micro\meter}, which assured
	that deformed areas do not overlap. A custom-written LabVIEW software (National Instruments) was used to process signals and to control the instrument through a data acquisition card
	(PCIe-6361, National Instruments).
	
	Ferrule-top probes of 0.2-0.5~N/m stiffness and 60-105~\SI{}{\micro\meter} bead radius were selected for these experiments and calibrated according to~\cite{beekmans_metrological_2015}. Two indentation-controlled profiles were selected for the characterization of depth and frequency dependent viscoelasticity: oscillatory ramp loading (OR) and equilibrium frequency sweep (FS). Fig.~\ref{profiles} shows the typical curves of the controlled-indentation and load response. Depth-controlled oscillatory ramp indentations (Fig.~\ref{profiles}a) had small 0.2~\SI{}{\micro\meter} oscillations at 5.62~Hz frequency superimposed on top of a
	loading ramp at 0.01 strain rate estimated by $\dot\varepsilon \sim \Delta\varepsilon/t$. Depth-controlled equilibrium frequency sweep measurements (Fig.~\ref{profiles}b) consisted of the
	loading part up to 10~\SI{}{\micro\meter} with 10~\SI{}{\micro\meter}/s indentation speed, followed by 30~s stress relaxation period to reach mechanical equilibrium and series of small ($h_{0} = 0.2$~\SI{}{\micro\meter}) sinusoidal
	oscillations at five distinct frequencies: 1, 1.78, 3.2, 5.62 and 10~Hz. The approach speed was set to 30~\SI{}{\micro\meter}/s, the surface of the sample was determined, and an indentation-controlled feedback was triggered at approximate load of 15~nN, which resulted in the
	initial uncontrolled 1-3~\SI{}{\micro\meter} indentation depth, which was later corrected in post processing procedures.
	
	\begin{figure}
		\centering
		\includegraphics[scale=0.60]{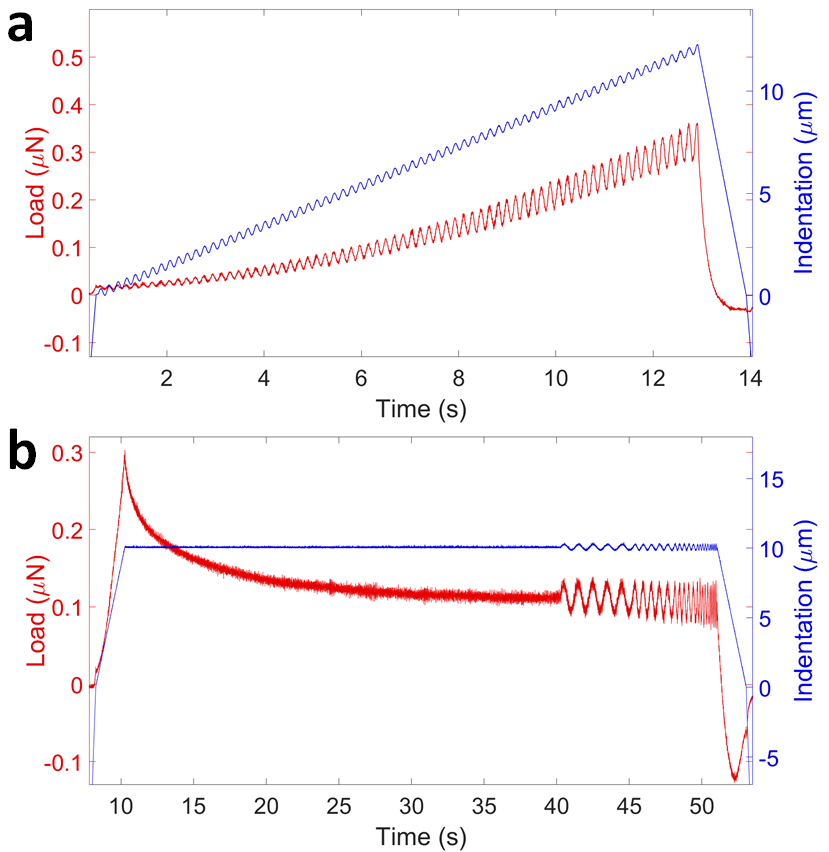}
		\caption{\label{profiles} Depth-controlled indentation profiles. (a) Oscillatory ramp loading profile at 5.62~Hz oscillation frequency and (b) equilibrium frequency sweep profile between 1-10~Hz frequency range with both oscillation amplitudes of 0.2~\SI{}{\micro\meter}. }
	\end{figure}

	\subsection*{Data analysis}
	\label{analysis}
	Raw data was analyzed with custom-written MATLAB functions. The Hertz model was used to fit an initial loading data up to the cantilever threshold value to obtain the true surface position:
	\begin{align}
		F=\dfrac{4}{3}\dfrac{E}{1-\nu^{2}}\dfrac{R}{\sqrt{h^{3}}},
	\end{align}
	
	\noindent where $F$ is the load, $E$ is an elastic modulus, $\nu$ is the Poisson's ratio of compressibility (we assume that brain is incompressible $\nu$ = 0.5), $h$ is the indentation depth. This allows us to correct the indentation depth $h$ and estimate the strain for measurements with probes of different radius:
	$\varepsilon = 0.2\times a / R$, where contact radius was estimated as $a = \sqrt{hR}$ varying between 22-39~\SI{}{\micro\meter}. Strain of 7.3$\%$ was selected for comparative analysis in order to
	fulfill small strain approximation $\varepsilon < 0.08$~\cite{lin_spherical_2009}.
	
	The sinusoidal oscillations were fit to cosine function, obtained amplitudes and phases were used to calculate the storage and loss moduli $E$\textasciiacute and $E$\textacutedbl~\cite{hoorn_local_2016}, which is a measure of elasticity and viscosity, respectively:
	
	\begin{align}
		\dfrac{E\text{\textasciiacute}(\omega)}{1-\nu^{2}}&=\frac{F_{0}}{h_{0}} cos\delta\dfrac{\sqrt{\pi}}{2}\dfrac{1}{\sqrt{A}}\label{eqn:storage},
			\end{align}
	
	\begin{align}
		\dfrac{E\text{\textacutedbl} (\omega)}{1-\nu^{2}}&=\frac{F_{0}}{h_{0}}sin\delta\dfrac{\sqrt{\pi}}{2}\dfrac{1}{\sqrt{A}}\label{eqn:loss}
	\end{align}
	
	\noindent where $\omega$ is the frequency, $F$$_{0}$ and $h$$_{0}$ are the amplitudes of oscillatory load and indentation depth, respectively, $\delta$ is the phase-shift between the recorded indentation and load oscillations, $A = \pi a^{2}$ is the contact area.
	
	The contact area changes with the depth during oscillatory ramp, thus every 5 cycles were used for fitting and averaged
	indentation depth was used for the calculation of $E$\textasciiacute and $E$\textacutedbl. Finally, all cosine fits with the $R^{2} \leq 0.7$ and measurements which started in contact
	were rejected.
	
	Normality of data distribution was tested with Shapiro-Wilk test ($n$ $\geq$ 3). In case of normal distribution, statistical differences between multiple groups were investigated with
	one-way ANOVA followed by Bonferroni \textit{post hoc} test to achieve 95$\%$ confidence level ($\alpha = 0.05$). For
	non-normally distributed data, Kruskal-Wallis ANOVA test followed by Dunn's \textit{post hoc} test with
	$\check{S}$id$\acute{a}$k correction for $\alpha_{1}$ = 1 - (1 - $\alpha$)$^{1/k}$ was used to compare multiple groups. All statistical analyses were performed with Statistics and Machine
	Learning Toolbox (version 2017a, The Mathworks, Natick, MA, USA).

	\subsection*{Code availability}
	The computer code used to generate the results of this study is available on request from the corresponding author.
	
	\subsection*{Data availability}
	All raw and processed data that support the findings of this study are available from the corresponding author upon request.
}
\section*{Acknowledgements}
The research leading to these results has received funding from the European Research Council under the European Union's Seventh Framework Programme (FP/2007–2013)/ERC grant
agreement no. [615170], the Dutch Technology Foundation (STW) under the iMIT program (P11-13) and Foundation for Fundamental Research on Matter (FOM), which is financially supported by the Netherlands Organisation for Scientific Research (NWO). The authors further thank M.Marrese, H.van Hoorn, T.Smit and L.Kooijman for fruitful discussions.
\section*{Author contributions}
N.A., W.J.W., E.M.H. and D.I. designed the research; S.V.B. and N.A. performed the preliminary experiments; N.A. performed the experiments and analyzed the data; N.A. and D.I. wrote the manuscript; all authors contributed to the editing of the paper.

\section*{Competing interests}
D.I. is co-founder and shareholder of Optics11.

\bibliography{MyLibrary}
\bibliographystyle{unsrt}
	
\end{document}